\documentclass[aps, singlecolumn, showpacs]{revtex4-2}
\usepackage{amsmath}
\begin{document}
\title {A note on diffeomorphism-invariance in gravitational field equations}

\author{S. C. Tiwari \\
Department of Physics, Institute of Science, Banaras Hindu University, Varanasi 221005, and \\ Institute of Natural Philosophy \\
Varanasi India\\ Email address: $vns\_sctiwari@yahoo.com$ \\}
\begin{abstract}
The notion of diffeomorphism invariance and general covariance are conceptually delicate issues for the field equations and the actions. A thorough study on the original Einstein field equation and its two modifications by Einstein is presented. It is concluded that the cosmological constant and the unimodular condition are independent concepts, and the unimodular condition has no direct role in the derivation of the traceless field equation. Unimodular gravity with unambiguous action does not exist, and the derivation of traceless equation from the action principle is an open question.
\end{abstract}

\maketitle

\section{\bf Introduction}

Mathematically diffeomorphism (Diff) is a $C^\infty$ map $\phi : M \rightarrow N$ that is one-to-one, onto, and its inverse is $C^\infty$. The manifolds $M$ and $N$ have same dimension and identical structure. If $T$ is a tensor field on $M$, $\phi$ is a symmetry transformation for the tensor $T$. For the metric tensor field $g$, Diff is called an isometry \cite{1,2}. How does one relate Diff with the physical space-time?  In the active viewpoint Diff for physical tensor fields in the 4D space-time manifold corresponds to the gauge freedom of general relativity \cite{1}. In this viewpoint, the coordinate systems are not considered, and $\phi$ is a map from tensors at $p\in M$ to tensors at $\phi (p)$. Alternatively, let us consider a coordinate system then $\phi$ is a map for the coordinate transformation $ x^\mu \rightarrow x^{\prime \mu}$ that leaves $p$ and all tensors at $p$ invariant. This is termed a passive viewpoint on Diff. The so-called passive viewpoint is essentially the one that led to the introduction of the Christoffel symbols and the Ricci calculus in the Riemannian geometry, see nice historical and insightful account on differential geometry by Chern in \cite{3}. On the other hand, the active viewpoint seems natural for the formalism based on differential forms and gravity as a gauge theory, see Regge in \cite{3}. A short exposition on the differential forms and frame field is given in Chapter 7 of the monograph \cite{4}. Authors explain that for tetrad or vierbein, it is better to use the term frame field. 

In the literature on gravity, general relativity (GR), and unimodular gravity (UG) Diff and general covariance have been discussed with various angles and perspectives. The discussions on Diff are related with the field equations and/or the actions. While discussing symmetry transformations and invarince one has to be careful that the invariance of the field equations under a symmetry transformation, in general, may not necessarily be an invariance of the action. Another important point is regarding the role of the symmetry transformation in the solutions of the field equations. Finally, one may need to distinguish between manifest invariance/covariance and the invariance in practice. 

Let us take the example of the classical electrodynamics (CED) to explain above points. Maxwell field equations for electric and magnetic fields $({\bf E}, {\bf B})$ are not only Lorentz invariant, they are invariant under a 15-parameter group of the conformal transformations \cite{5}. One can re-write them in manifestly Lorentz-covariant form introducing the electromagnetic field tensor $F_{\mu\nu}$. They are gauge-invariant under the transformation $A_\mu \rightarrow A_\mu + \partial_\mu \chi $ since $F_{\mu\nu}= \partial_\mu A_\nu - \partial_\nu A_\mu$.  They are also invariant under duality rotation \cite{4}. The form of the Maxwell field equations has simple transcription to general-covariant form making use of the metric tensor $g_{\mu\nu}$ in the pseudo-Riemannian space-time. The action for CED too respects these symmetries. However, under local duality rotation the generalized Maxwell field equations derived in \cite{6} are invariant whereas the action is not. Regarding the gauge symmetry, choosing a gauge condition while seeking the solution of the field equations restricts gauge freedom. Moreover, in the action the variational principle is based on the electromagnetic potentials $A_\mu$ as independent field variables, therefore, the question of gauge-invariance becomes subtle; more so when one seeks quantization of the fields \cite{7}. Even in CED, the conserved quantities like the canonical energy-momentum tensor and canonical angular momentum tensor are not gauge-invariant. It is clear that to comprehend the whole picture of symmetry and invariance one has to consider field equations, action, and the solutions.

In most of the literature the condition on the determinant of the metric
\begin{equation} 
\sqrt{-g} =1 
\end{equation}
defines UG, and the issue of Diff is addressed based on it. In a recent paper, a thorough study on UG \cite{8} underlines three points: (A) the assumption (1) on the determinant does not mean discarding general covariance, (B) UG needs to be viewed independently of the cosmological constant problem, and (C) Unimodular Relativity (UR) is defined by equi-projective geometry and condition (1) on the fundamental metric tensor $f_{\mu\nu}$. The present note makes a systematic study on Diff taking into account the complete picture on the solutions and the field equations in the light of the insights gained in \cite{8}. In the next section, Einstein-Eddington arguments on the assumption (1), the Cauchy problem and unique solution of the Einstein field equations \cite{2}, the role of (1) in the Schwarzschild solution \cite{9} and Kerr-Schild metric \cite{10} are discussed. Section III revisits the Einstein field equation \cite{11} and the modified field equation with a cosmological constant \cite{12}. The traceless field equation \cite{13} linked to UG in the literature is discussed in Section IV.  The last section concludes with remarks on classical general relativistic and particle physics-oriented paradigms.

\section{\bf Unimodular condition}

A brief comment on the interpretation of the condition (1) following Einstein \cite{11} and Eddington \cite{14} has been made in \cite{8}. However, it needs elaboration to bring clarity on the physical import of the unimodular condition. Einstein considers the transformation of the determinant $\sqrt{-g}$ and the volume element $dV$ under the coordinate transformation $x^\mu \rightarrow x^{\prime \mu}$ to establish that $\sqrt{-g}~ dV$ is invariant. The magnitude of this quantity is measured in a local frame of reference with rigid rods and clocks similar to what is done in special relativity. If $\sqrt{-g}$ is positive and finite then one can choose a coordinate system "a posteriori" such that it is always unity. This choice is merely a calculational tool for convenience, and the general covariance can be restored later. Eddington elaborating on it notes that one can choose $\sqrt{-g}=1$ in any space-time; tensor densities now become tensors. 

I think in Einstein's approach the role of the condition (1) in the Einstein field equations is analogous to the role of EM potentials in CED treated as calculational tools having no physical reality. It is well known that after the experimental proof of the Aharonov-Bohm effect there is a strong argument for elevating the status of the EM potentials to fundamental physical fields. In the same spirit, one could raise the assumption (1) to a principle signifying new law of gravity (UG). 

The role of Diff invariance in the Cauchy problem for the Einstein field equations needs careful technical considerations \cite{2}. Here we note the important point that the solutions of the field equations are unique only up to a Diff. Therefore, it becomes necessary to restrict Diff using some kind of gauge conditions. Assuming a background metric, the gauge conditions are defined in terms of the covariant derivatives of the physical metric with respect to the background metric. An interesting example is that of the harmonic gauge. In the literature, one usually draws analogy to the Lorentz gauge condition in CED.  

The significance of the unimodular assumption (1) in the standard solutions of the field equations is quite interesting. In the original derivation \cite{9} Schwarzschild used a privilaged coordinate system $(t, x, \psi, \phi)$ defined in terms of the spherical coordinates $(x=\frac{r^3}{3}, ~\psi =-\cos \theta, ~\phi)$ with the line element in the flat space-time
\begin{equation}
ds^2 =-dt^2 +\frac{dx^2}{r^4} +r^2(\frac{d\psi^2}{\sin^2\theta} +\sin^2\theta ~d\phi^2)
\end{equation}
It is easy to verify that the determinant of this metric is equal to -1. To solve the Einstein field equations in vacuum the static spherically symmetric metric (2) is generalized to
\begin{equation}
ds^2 =-f_0 ~dt^2 +f_1~dx^2 +f_2~(\frac{d\psi^2}{\sin^2\theta} +\sin^2\theta ~d\phi^2)
\end{equation}
and for preserving unimodular condition the functions are set to satisfy $f_0(x) f_1(x)f_2^2(x) =1$. Various steps to reach the final form of the Schwarzscild metric can be found in \cite{9}. 

The Kerr-Schild form of the metric is an important class of the exact solutions of the Einstein field equations. It is given by
\begin{equation}
g_{\mu\nu}= \eta_{\mu\nu} -2V ~k_\mu k_\nu
\end{equation}
Here $\eta_{\mu\nu}$ is a flat space-time metric, $V$ is a scalar function, and $k_\mu$ is a null vector with respect to $\eta_{\mu\nu}$
\begin{equation}
k^\mu k_\mu=0
\end{equation}
It can be verified that $k_\mu$ is a null vector also with respect to $g_{\mu\nu}$. The main point relevant for the present discussion is that the determinant of this metric satisfies $\sqrt{-g} =1$ \cite{10}. 

\section{\bf Einstein's field equation and modification}

The foundations of the famous gravitational field equation were discussed by Einstein himself in \cite{11}. Later, Einstein modified the field equation based on the cosmological considerations \cite{12}. The trace-free field equation was proposed in 1919 by him seeking the role of gravity in the structure of the electron \cite{13}. We revisit the three field equations to gain new insights on the nature of the relation between geometry and matter/field. Trace-free field equation is discussed in the next section.

The gravitational field equation equates the geometric quantity, the Einstein tensor in 4D pseudo-Reimannian space-time, $G_{\mu\nu}$ with the energy-momentum tensor $T_{\mu\nu}$ of the matter and physical fields. The Einstein tensor is defined to be
\begin{equation}
G_{\mu\nu} = R_{\mu\nu} -\frac{1}{2} g_{\mu\nu} R
\end{equation}
and the Einstein field equation is
\begin{equation}
G_{\mu\nu} =\alpha_G T_{\mu\nu}
\end{equation}
Here the gravitational coupling constant is $\alpha_G = \frac{8 \pi G}{c^4}$ where $G$ is the Newtonian gravitational constant, and $c$ is the velocity of light in vacuum. The textbooks invoke the covariant divergence law for $T_{\mu\nu}$ and the Bianchi identity satisfied by the Einstein tensor for logical justification of the Einstein field equation (7). 

However, the role of the physical principle of equivalence of Einstein \cite{11} needs careful attention. Einstein in section(2) of \cite{11} considers uniform acceleration of a frame of reference $K^\prime$ with respect to a frame $K$ in a Galilean system of reference, and argues that general relativity leads to a gravity field by changing the frame of reference. Eddington in section (17) of \cite{14} presents a nice critique on the equivalence principle terming it a hypothesis. In section (54) he gives a new derivation of the Einstein field equation (7). 

Note that the first mathematical step in Einstein's derivation of the field equation is  that of postulating the equation of motion of a point mass 
as a geodetic line in the curved space-time
\begin{equation}
\frac{d^2x^\mu}{ds^2} + \Gamma^\mu_{~\nu\sigma} \frac{dx^\nu}{ds} \frac{dx^\sigma}{ds}=0
\end{equation}
Einstein argues that if $\Gamma^\mu_{~\nu\sigma} $ vanishes then the geodetic line defined by Eq.(8) reduces to a straight line uniform motion, therefore, the Christoffel symbols represent the gravitational field. The next step is to generalize the energy-momentum conservation law in the Galilean coordinates to the general coordinates system in the form of the covariant divergence law
\begin{equation}
\frac{\partial T^{\mu\nu}}{\partial x_\nu}=0  \Rightarrow T^{\mu\nu}_{~:\nu} =0
\end{equation}
The equivalence principle is embodied in the hypothesis (9). Eddington argues that this hypothesis is unnecessary, and postulates that the gravitational field equation is given by (7). In that case, the covariant divergence law for $T_{\mu\nu}$ becomes an identity. Thus, rather than the principle of equivalence it is the principle of identification in the deductive approach of Eddington.

Einstein was not satisfied with the right hand side of the field equation (7), and believed that it had a provisional role. Unless one has the equation for the total field, GR is an incomplete theory \cite{15}. In one of the modern approaches Bondi \cite{16} too is critical of the role of the equivalence principle in GR. His argument is that gravitation is observable because the acceleration varies from place to place, therefore, the acceleration 4-vector $f_\mu$ could be postulated to satisfy the geodesic deviation law 
\begin{equation}
\delta f^\mu = R^\mu_{~\nu\lambda\sigma} v^\lambda v^\sigma \delta x^\nu
\end{equation}

For a brief review and critical assessment on this issue we refer to \cite{17}. The present discussion shows that there is no uniqueness in the form of the gravitational field equation of Einstein (7). Even the argument of logical simplicity is not absolute, for example, the first-order formulation of the field equation may be viewed simpler than (7). Keeping Diff invariance intact one may seek modifications of the Einstein field equation (7). A modifications motivated by cosmological considerations was suggested by Einstein himself. 

In 1917 Einstein proposed the following field equation replacing Eq. (7)
\begin{equation}
G_{\mu\nu} -\lambda g_{\mu\nu} = \alpha_G T_{\mu\nu}
\end{equation}
Here $\lambda$ is a new universal constant, known as the cosmological constant. Note that the modified field equation (11) is Diff invariant.  
Later on Einstein wrote \cite{18} that, 'The introduction of the second member constitutes a complication of the theory, which reduces its logical simplicity'. Contrary to this, revisiting the arguments put forward by Einstein to propose this modification \cite{12} it emerges that he gave sound logical foundation for Eq.(11) that could be viewed as a generalized field equation that includes the Einstein field equation (7) as a special case setting $\lambda =0$. 

Einstein observes \cite{12} that there is a crucial difference between the problem of the planetary system and that of the Universe treated in GR. The boundary conditions at spatial infinity become complicated. In fact, the Newtonian theory also suffers from these difficulties. To remedy it, the Poisson equation for the gravitational potential is modified
\begin{equation}
\nabla^2 \Phi -\lambda \Phi = \frac{\alpha_G c^4}{2} \rho
\end{equation}
where $\rho$ is mass density. Just as the Einstein field equation is in a sense a GR transformation of the Poisson equation, the modified equation (11) is a logical GR extension of the modified Poisson equation (12). A lucid discussion on the modified Poisson equation (12) and a short review on the cosmological constant can be found in \cite{19}.

In the original derivation of Eq.(7) as well as that of Eq.(11) Einstein based his arguments on the Newtonian theory and the physical properties of the matter. We have pointed out that Eddington postulated the field equation (7) dispensing with the equivalence principle. In the later exposition \cite{18} Einstein invoked three conditions to derive the field equation (7) as follows. It is assumed that the Poisson equation for gravitational potential serves the guiding principle for the nature of the field equation for the metric tensor $g_{\mu\nu}$ interpreted as gravitational potential in GR.  First two conditions are (1) differential coefficient of the metric tensor higher than the second should not occur, and (2) the field equation should be linear in these second-order differentials. The third condition is that the covariant divergence of this geometric quantity should vanish to respect the covariant divergence law for the energy-momentum tensor of matter. Now, from the Riemann curvature tensor the geometric quantity constructed based on the first two conditions is $R_{\mu\nu} +a g_{\mu\nu} R$, and the third condition fixes $a=-\frac{1}{2}$. If Einstein's prescription is adopted using the modified Poisson equation (12) then the geometric quantity should include a linear term in the potential, i. e. $g_{\mu\nu}$, therefore, one has $R_{\mu\nu} +a g_{\mu\nu} R +\lambda g_{\mu\nu}$ leading to the modified Einstein field equation (11).

Departing from Einstein-Eddington approaches we suggest an alternative derivation based purely on the geometric identity of the pseudo-Riemannian 4D space-time. A lengthy proof of the identity
\begin{equation}                         
G^{\mu\nu}_{~:\nu} =0
\end{equation}
is given in section (52) by Eddington \cite{14}. In section (61) the invariance of an invariant function of $g_{\mu\nu}$ under the general coordinate transformations is shown to lead to a covariant divergence  law for the Hamiltonian derivative of the invariant. In particular, the divergence law (13) follows if the invariant is Ricci scalar $R$ for which the Hamiltonian derivative is the Einstein tensor $G_{\mu\nu}$. The Bianchi identity (13) is independent of any physical arguments or any postulate on the field equation. Therefore, beginning with Eq.(13) we seek its solution, and equate it with matter/field energy-momentum tensor that obeys the covariant divergence law (9). The general solution of (13) is evidently $G_{\mu\nu} +\lambda g_{\mu\nu}$ resulting into the modified field equation with a cosmological constant that arises as an integration constant remembering that the covariant divergence of $g_{\mu\nu}$ is zero. Thus, the cosmological constant is not an ad hoc universal constant the way Einstein introduced it. 

\section{\bf Traceless field equation}

In 1919 Einstein asked the question if gravitation had any role in the structure of the electron \cite{13}. The field equation (7) with the symmetric energy-momentum tensor for the EM field $E_{\mu\nu}$ is examined: it is shown that the current-density vanishes everywhere. A new trace-free field equation is proposed
\begin{equation}
R_{\mu\nu} -\frac{1}{4} g_{\mu\nu} R = \alpha_G E_{\mu\nu}
\end{equation}
Since $g^{\mu\nu} E_{\mu\nu} =0$, Eq.(14) is an identity $0=0$. The trace-free equation is generalized to any matter tensor energy-momentum tensor $T_{\mu\nu}$
\begin{equation}
R_{\mu\nu} -\frac{1}{4} g_{\mu\nu} R= \alpha_G (T_{\mu\nu} -\frac{1}{4} T g_{\mu\nu})
\end{equation}

The conceptual problems related with the traceless equation (15) arise for two reasons: (1) Einstein in his paper \cite{13} revisited the modified field equation (11) in the light of the traceless equation and established that the cosmological constant was not a universal constant but an integration constant, and (2) though Einstein did not use the action principle, Anderson and Finkelstein \cite{20} addressed the question of the cosmological constant based on an action principle postulating the unimodular condition and derived the traceless equation in unimodular gravity. The advances in the observational cosmology and the importance of the physical interpretation of the cosmological constant, as well as the advances in astro-particle physics revived great interest in both modified Einstein field equation and UG. Unfortunately, many authors make erroneous statement that UG is due to Einstein, and that the traceless field equation is not Diff invariant. To clarify the misconceptions and to gain new physical insights we present a thorough discussion in this section.

{\bf Cosmological constant : ~} In the Appendix of \cite{8} we have given the main steps that Einstein used to arrive at Eq.(15), and to prove that the cosmological constant was an integration constant. Einstein does not offer any reason as to why the cosmological constant for the Universe at large should be relevant for the structure of an elementary particle. Not only this, the electron problem remains unsolved. Since all the derivations in \cite{12, 13} respect Diff invariance, the traceless field equation respects Diff invariance. Einstein himself does not use the unimodular condition any where to get the field equation (15), therefore, as pointed out in Section II,  the use of Eq.(1) in Einstein's work is merely that of a calculational tool. Therefore, it is historically and conceptually wrong to associate Einstein with UG. 

The cosmological constant and the unimodular condition are independent physical concepts. It is not recognized in most of the current literature that the cosmological constant in the gravitational field equation arises in the unified theory of Weyl (1918) in which the action is assumed to be the square of the Ricci scalar in Weyl geometry \cite{21}. Even in the Weyl-Dirac theory \cite{22} the cosmological constant is present in the field equation. However, the remarkable point made by Dirac \cite{22} is that, 'It appears as one of the fundamental principles of Nature that the equations expressing basic laws should be invariant under the widest possible group of transformations'. In Weyl geometry, the wider group of transformation is that of general coordinates (Diff) and the gauge transformations. In a theory that has this wider group of transformations the cosmological constant arises inevitably. In contrast, UG has restricted group of transformations. Logically, one expects to dissociate cosmological constant from UG or rather view UG and cosmological constant as independent concepts \cite{8}.

Let us consider the formulation based on the action principle. Anderson and Finkelstein \cite{20} in the abstract of their paper state that the cosmological constant "appears as an inelegant ambiguity " in the action in general relativity. Contrary to this statement we argue that the construction of the action based on the invariants leads to the simplest and natural combination comprising of $\sqrt{-g}$ and $R$. The action containing only $\sqrt{-g}$ is not interesting and physically vacuous since the variation in the action gives the field equation $g_{\mu\nu} =0$. Note that the invariance under general coordinate transformations of the Hamiltonian derivative  for this invariant leads to $g^{\mu\nu}_{~:\nu}=0$. Now, the action could be assumed to be
\begin{equation}
S_\lambda= \int (R \sqrt{-g}+\lambda \sqrt{-g}) dV 
\end{equation}
Using the variational principle taking $g_{\mu\nu}$ as the independent field variable leads to the modified field equation with a cosmological constant. Here
\begin{equation}
d(\sqrt{-g}) =\frac{1}{2} \sqrt{-g} g^{\mu\nu} d g_{\mu\nu}
\end{equation}

Authors in \cite{20} make important observations on the reducibility of the metric tensor to the absolute value of the determinant of the metric tensor and the relative tensor of determinant equal to -1 based on physical measurements and light-cone structure. Eq.(1) is a constraint equation in their action formulation and the cosmological constant is introduced as a Lagrange multiplier. Remarkably, the action is Diff invariant, and the traceless field equations do not follow directly from the variation of the action. The field equation obtained from the variations in $g_{\mu\nu}$ is in the form of the modified field equation (11) in vacuum ($T_{\mu\nu} =0$). Variation in $\lambda$ gives the constraint (1). Taking the trace of the vacuum field equation (11) determines the Lagrange multiplier to be equal to $-\frac{1}{4} R$, and the field equation transforms to
\begin{equation}
R_{\mu\nu} -\frac{1}{4} R g_{\mu\nu}=0
\end{equation}
This formulation, in view of the role of the constraint (1) is termed as UG. The authors rightly note that unimodular action "is still generally covariant". Using the Bianchi identity (13) it follows that the cosmological constant is an integration constant. The field equation in the presence of the matter energy-momentum tensor for the UG (Equation (7) in \cite{20}) is obtained using similar steps. However, in this case, the cosmological constant is not necessarily an integration constant as the additional assumption on the covariant divergence law could be relaxed \cite{23}.

If we carefully examine the derivation of the traceless equation (18) we find that the unimodular condition (1) does not play any role in this derivation. Only the structure of the UG action in which $\lambda$ is a Lagrange multiplier as compared to the action (16) in which it is a constant is different. As far as the derivation of the traceless equation is concerned exactly the same steps are required in both cases. In fact, much stronger argument to de-link unimodular condition from the traceless equation can be made. The expression for the Ricci tensor is
\begin{equation}
R_{\mu\nu} = \Gamma^\alpha_{~\mu\sigma} \Gamma^\sigma_{~\alpha \nu} -\Gamma^\alpha _{~\mu\nu} \Gamma^\sigma_{~\alpha\sigma} + \frac{\partial \Gamma^\sigma_{~\mu\sigma} }{\partial x^\nu} -  \frac{\partial \Gamma^\sigma_{~\mu\nu} }{\partial x^\sigma}
\end{equation}
Setting $ \sqrt{-g}=1$ in the expression (19) we get for UG Ricci tensor
\begin{equation}
\bar{R}_{\mu\nu} =\Gamma^\alpha_{~\mu\sigma} \Gamma^\sigma_{~\alpha \nu}-  \frac{\partial \Gamma^\sigma_{~\mu\nu} }{\partial x^\sigma}
\end{equation}
and $\bar{R} = g^{\mu\nu}  \bar{R}_{\mu\nu}$. It is obvious that the Einstein field equation in vacuum becomes
\begin{equation}
\bar{R}_{\mu\nu} -\frac{1}{2} \bar{R} g_{\mu\nu}=0
\end{equation}
and it is not equivalent to the traceless equation (18) and cannot to be transformed to the  tracelss form.

{\bf Electron Structure:~} Discovery of electron in 1897 by J. J. Thomson associates particulate nature to the atom of electricity having physical attributes of mass and charge. Is electron a point charge? Does electron mass originate from the electromagnetic fields? Does electron have extended spatial structure? Eminent physicists Thomson, Poincare, Lorentz, Abraham, Mie and Weyl were deeply involved in these fundamental questions. Einstein \cite{13} does not seem to realize the import of these questions, and mentions rather superficially the efforts of Mie and Weyl. Weyl \cite{21} in Section 26 gives a thorough discussion on Mie's theory. In Section 32, the electron model based on the Einstein-Maxwell theory is discussed by him. The problem of electron structure in his unified theory of gravitation and electromagnetism is presented in Secion 36. 

Einstein in 1935 with Rosen \cite{24} returned to the electron problem, and once again sought modification of the Einstein field equation. In the Abstract the authors write:"By the considerations of a simple example they are led to modify slightly the gravitational equations which then admit regular solutions for the static spherically symmetric case." Taking a simple example of the metric tensor, it is shown that the determinant of $g_{\mu\nu}$ vanishes at some point in space, therefore, $g^{\mu\nu}$ becomes infinite. The Riemann curvature tensor and the Ricci tensor become indeterminate $\frac{0}{0}$. The modified field equation (Eq.(3a) in \cite{24}) is $R^*_{\mu\nu} = g^2 R_{\mu\nu} =0$. Instead of the usual contravariant metric tensor $g^{\mu\nu}$, the cofactors of $g_{\mu\nu}$ in $g$ are introduced.   The concept of Einstein-Rosen bridge in this paper has attracted great attention during past few years in connection with the speculation ER=EPR. However, the electron problem continues to remain unresolved; it has become much more intricate as electron has additional physical attributes of spin and anomalous magnetic moment \cite{25}. Thus, the traceless field equation has lost any relevance for the original electron problem envisaged by Einstein \cite{13}.

\section{\bf Conclusion}

A thorough study presented in this paper on the foundations of UG establishes that (1) the cosmological constant and the unimodular condition are independent issues, (2) the traceless field equation is Diff invariant, (3) there is no direct role of the unimodular condition in the derivation of the traceless equation, and (4) the derivation of the traceless equation directly from the action principle is an open question.

The Diff invariance in the field equations and the action principles needs further investigation \cite{26}.  Regarding the cosmological constant and the dark energy inspired by the observed accelerating Universe, an alternative seemingly conservative but radical approach based on \cite{28} deserves attention; a more radical approach could be based on \cite{17}.

\end{document}